-----------------------------------------------------------------------

\documentstyle[12pt]{article}
\begin{document}

\title{The 2+1 Dirac equation with the $\delta$ potential}

\author{Shi-Hai Dong $^{1, 2, 3}$\thanks{E-mail address: dongsh@nuclecu. unam. mx} and Zhong-Qi Ma $^{1}$\\
{\footnotesize $^{1}$ IHEP, P. O. Box 918(4),
Beijing 100039, People's Republic of China}\\
{\footnotesize $^{2}$
Department of Physics, Kansas State University, Manhattan 66506, USA}\\
{\footnotesize $^{3}$
Instituto de Ciencias Nucleares, UNAM, A. P. 70-543, Circuito Exterior, C. U. , 04510 Mexico}}

\date{}

\maketitle

\begin{abstract}
In this study the bound state of 2+1 Dirac equation with the
cylindrically symmetric $\delta (r-r_{0})$-potential
is presented. It is shown that the energy can be obtained by the
transcendental equation obtained from the matching condition
in the configuration space. It is surprisingly found that the relation between the radial functions $F_{jE}^{+}$, $G_{jE}^{+}$ and $F_{jE}^{-}$, $G_{jE}^{-}$ can be established by $SO(2)$ group.

\end{abstract}


\newpage
\vskip 0.5cm
\begin{center}
{\large\bf {1. Introduction}}
\end{center}

As we know, it is a curious and complex situation to solve
the Dirac equation with the
$\delta $ potential in comparison with the equivalent problem
in non-relativistic quantum mechanics,
i. e. , the Schr\"{o}dinger equation,
which is discussed in any course on quantum mechanics.
This is basically related to the fact that being the Dirac equation of
first order, a singular potential, like the $\delta $ one,
induces discontinuities at the level of the wave function themselves
instead of the usual discontinuities that appear
in the first derivative in the Schr\"{o}dinger equation.

This puzzling situation has been discussed previously [1-2].
The partial wave operators have been constructed by making use of the self-adjoin
extension theory. Recently, other authors have shown how to deal with the Dirac
equation with the $\delta$-potential in the different spaces [3-6]. The
position-space treatment of this problem has been carried out in [3, 4]. However, it has been discussed in the momentum space [5] as well as in
the configuration-space treatment [6].
All of those papers mentioned above have mainly discussed this problem
in 3+1 dimensional space.
The study of this similar problem in one dimension has
been completed in [7, 8].
On the other hand, the positron emission which is
related with this problem has been studied in Ref. [9].

With the interest of the lower-dimensional field theory, it is necessary to
study the 2+1 Dirac equation with the $\delta$-potential,
which has never been addressed in the literature, to our knowledge.
The purpose of this paper is to study this problem in the configuration
space.

This paper is organized as follows. Section 2 is devoted to the introduction of
the 2+1 Dirac equation. The bound state will be discussed
in Section 3. A conclusion will be given in Section 4.

\vskip 0.5cm
\begin{center}
{\large\bf {2. Dirac equation in 2+1 dimensions}}
\end{center}

It is convenient to recall some general properties of the solution
of the Dirac equation in a central potential.
For more details the reader may consult
the book [10].

For the present discussion we do not need the explicit form of the
angular part. Consider the 2+1 Dirac equation

$$\displaystyle \sum_{\mu=0}^{2}i\gamma^{\mu}
\left(\partial_{\mu}+ieA_{\mu}\right)\psi=M\psi
\eqno (1) $$
where $M$ is the mass of the particle, and
$$\gamma^{0}=\sigma_{3}~~~~~
\gamma^{1}=i\sigma_{1}~~~~~\gamma^{2}=i\sigma_{2}. \eqno (2) $$

\noindent
Throughout this paper the natural units $\hbar=c=1$ are employed if not
explicitly stated otherwise.
Discuss the special case where only the zero component of
$A_{\mu}$ is non-vanishing and cylindrically symmetric:
$$A_{1}=A_{2}=0~~~~~eA_{0}=V(r) \eqno (3) $$
where $V(r)$ is taken as the attractive
cylindrically symmetric potential
$$V(r) = -a \delta (r-r_{0})~~~a>0\eqno(4)$$
Let
$$\psi_{jE}(t, {\bf r})=e^{-iEt}r^{-1/2}\left(\begin{array}{c}
F_{jE}(r)e^{i(j-1/2)\varphi} \\
G_{jE}(r)e^{i(j+1/2)\varphi} \end{array} \right), \eqno (5) $$
where $j$ denotes the total angular momentum, $j=\pm 1/2$,
$\pm 3/2$, $\ldots$. It is shown that the radial components
$F_{jE}$ and $G_{jE}$ satisfy the following set of coupled differential
equations [11]
$$\displaystyle {d \over dr}G_{jE}(r)+\displaystyle
{j \over r} G_{jE}(r)
=\left(E-V(r)-M\right)F_{jE}(r) , \eqno(6a) $$
$$-\displaystyle {d \over dr}F_{jE}(r)+\displaystyle
{j \over r} F_{jE}(r)
=\left(E-V(r)+M\right)G_{jE}(r) .  \eqno (6b) $$
It is easy to see that the
solutions with a negative $j$ can be
obtained from those with a positive $j$ by interchanging
$F_{jE}(r) \longleftrightarrow G_{-j-E}(r)$,
so that in the following we only discuss the solutions with
a positive $j$.

The physically admissible solutions are finite, continuous,
vanishing at the origin, and square integrable:
$$F_{jE}(r)=G_{jE}(r)=0, ~~~~~
{\rm when}~~r\longrightarrow 0, \eqno (7) $$
$$\displaystyle \int_{0}^{\infty} dr\left\{|F_{jE}(r)|^{2}
+|G_{jE}(r)|^{2}\right\} < \infty. \eqno (8) $$
The solutions for $|E|>M$ describe the scattering states, and
those for $|E|\leq M$ describe the bound states. In this note,
what we are interested in is related with the bound state.

Due to the linear discontinuity of the spinor function given by the
$\delta$-potential, we only
need to fix the boundary conditions in the neighborhood
of the shell $r=r_0$. Multiplying Eq. (6a) by $F_{jE}$ and Eq. (6b) by
$G_{jE}$ and calculating their summation, one can obtain
$$F_{jE}^{\prime} F_{jE}+G_{jE}^{\prime}G_{jE}=
-2MF_{jE}G_{jE} +j\frac{(F_{jE}^2-G_{jE}^2)}{r}\eqno(9)$$
where and hereafter we denote $F_{jE}(r)$ and
$G_{jE}(r)$ by $F_{jE}$ and $G_{jE}$ respectively and
the primes denote the radial derivatives with respect to the variable
$r$. Integrating between
$r_0-\varepsilon $ and $r_0+\varepsilon $ and taking
the limit $\varepsilon \rightarrow 0$ we get
$$\lim_{\varepsilon \rightarrow 0}
\int_{r_0-\varepsilon }^{r_0+\varepsilon}( F_{jE}^{\prime}F_{jE}+
G_{jE}^{\prime}G_{jE}) dr=
\lim_{\varepsilon \rightarrow0}\int_{r_0-\varepsilon }^{r_0+\varepsilon }
\left(-2MF_{jE}G_{jE} +j \frac{( F_{jE}^2-G_{jE}^2)}{r}\right)dr, \eqno(10)$$
which implies that
$$\left. \displaystyle \lim_{\varepsilon \rightarrow 0}( F_{jE}^2+G_{jE}^2)
\right| _{r_0-\varepsilon}^{r_0+\varepsilon }=0. \eqno(11)$$
It is found that the radial functions $F_{jE}$ and $G_{jE}$
can be regarded as the real and imaginary parts of a function in
$\rm C \negthinspace \negthinspace \negthinspace  \negthinspace 1$  .
From the geometrical point of view, it is found that the norm of the two
components spinor $F$ and $G$ are constant when crossing the support of
the $\delta$-potential, which coincides with the
results established in [1, 2]. In this way,
one can write the following boundary condition for all finite $r$
$${F^{+}_{jE}}^2+{G^{+}_{jE}}^2={F^{-}_{jE}}^2+{G^{-}_{jE}}^2. \eqno(12)$$
where $F^{\pm}_{jE}\equiv F(r_{0}\pm \epsilon)$ and
$G^{\pm}_{jE}\equiv F(r_{0}\pm \epsilon)$.
Let us now multiply Eq. (6a) by $G_{jE}$ and Eq. (6b) by $F_{jE}$
respectively and calculate their difference,
$$F_{jE}^{\prime}G_{jE}-F_{jE}G_{jE}^{\prime}
=-(E-M)G_{jE}^2+(E+M)F_{jE}^2+\frac{2jG_{jE}F_{jE}}{r}
-a\delta (r-r_0)( F_{jE}^2+G_{jE}^2)\eqno(13)$$
which is divided by $F_{jE}^2+G_{jE}^2$, one can integrate
in the neighborhood of the shell radius
$$\lim_{\varepsilon \rightarrow 0}\int_{r_0-\varepsilon }^{r_0+\varepsilon }
\frac{ F_{jE}^{\prime}G_{jE}-F_{jE} G_{jE}^{\prime}}{(F_{jE}^2+G_{jE}^2) }dr=
-a\lim_{\varepsilon \rightarrow 0}\int_{r_0-\varepsilon }^{r_0+\varepsilon}
\delta \left( r-r_0\right) dr. \eqno(14)$$
By using
$$\frac{F_{jE}^{\prime}G_{jE}-F_{jE} G_{jE}^{\prime}}{(F_{jE}^2+G_{jE}^2)}=
\left(\frac{F_{jE}}{G_{jE}}\right)^{\prime}\frac{1}{(F_{jE}/G_{jE})^{2}+1}
\eqno(15)$$
and since
$$\int {\frac 1{1+h^2(x)}d[h(x)]}=\arctan (h(x))$$
we have
$$\left. \displaystyle \lim_{\varepsilon \rightarrow 0}
\left(\arctan \frac{F_{jE}}{G_{jE}}\right)
\right|_{r_0-\varepsilon }^{r_0+\varepsilon }=-a. \eqno(16)$$
In this way, this boundary condition can be written as
$$\arctan \frac{F^{+}_{jE}}{G^{+}_{jE}}-\arctan \frac{F^{-}_{jE}}{G^{-}_{jE}}
=-a. \eqno(17)$$
Define the dimensionless parameter
$\alpha \equiv \tan(a)$, equation (17) can then be expressed as
$$\frac{F^{+}_{jE}}{G^{+}_{jE}}=\frac{(F^{-}_{jE}/G^{-}_{jE})
-\alpha}{1+ \alpha (F^{-}_{jE}/G^{-}_{jE})}. \eqno(18)$$
Except for an arbitrary phase, the last expression can
be written as a matrix relation between the radial functions
at both sides of the potential,
$$\left[\begin{array}{cc} F^{+}_{jE} \\ G^{+}_{jE} \end{array} \right]
=\left[
\begin{array}{cc}
\cos (a)  & -\sin (a) \\
\sin (a)  & \cos (a)
\end{array}
\right]
\left[\begin{array}{cc} F^{-}_{jE} \\ G^{-}_{jE} \end{array} \right]
\equiv A
\left[\begin{array}{cc} F^{-}_{jE} \\ G^{-}_{jE} \end{array} \right]\eqno(19)$$
where
$$A=\left[
\begin{array}{cc}
\cos (a)  & -\sin (a) \\
\sin (a)  & \cos (a)
\end{array}
\right]. $$

We note that $A$ is unitary and orthogonal, i. e. , $\det A=1$ and
contains the information for finding
the eigenvalue equation for the bound states. On the other hand, we see that matrix $A$ can construct the $SO(2)$ group, that is to say, the relation between the radial functions $F_{jE}^{+}$, $G_{jE}^{+}$ and $F_{jE}^{-}$, $G_{jE}^{-}$ at the both sides of the potential can be established by $SO(2)$ group, which is a new point, to our knowledge.
However, for the complex valued function with real and imaginary part
given by $F_{jE}$ and $G_{jE}$ respectively, it
is found that the $\delta $ function manifests
itself by a phase change of this function expressed by $\tan (a)$.

\vskip 0.5cm
\begin{center}
{\large\bf {3. Bound states}}
\end{center}

We now solve Eq. (6) for the energy $|E|\leq M$. In the region
$[0, r_{0}]$, we have
$$F^{-}_{jE}=e^{-i(j-1/2)\pi/2}\left[(M+E)\pi \kappa r/2\right]^{1/2}
J_{j-1/2}(i\kappa r) $$
$$G^{-}_{jE}=e^{-i(j-3/2)\pi/2}\left[(M-E)\pi \kappa r/2\right]^{1/2}
J_{j+1/2}(i\kappa r) \eqno (20)  $$
where $J_{m}(x)$ is the Bessel function, and
$$\kappa=\left(M^{2}-E^{2}\right)^{1/2}. \eqno (21) $$

\noindent
The ratio at $r=r_{0}-$ when $\lambda=0$ is
$$\left. \displaystyle {F^{-}_{jE} \over G^{-}_{jE}}
\right|_{r=r_{0}-}=-i \left(\displaystyle {M+E \over M-E }\right)^{1/2}
 \displaystyle {J_{j-1/2}(i\kappa r_{0}) \over J_{j+1/2}(i\kappa r_{0}) }
\eqno (22) $$

In the region $[r_{0}, \infty)$, we have $V(r)=0$ and
$$F^{+}_{jE}=e^{i(j+1/2)\pi/2}\left[(M+E)\pi \kappa r/2\right]^{1/2}
H^{(1)}_{j-1/2}(i\kappa r) $$
$$G^{+}_{jE}=e^{i(j+3/2)\pi/2}\left[(M-E)\pi \kappa r/2\right]^{1/2}
H^{(1)}_{j+1/2}(i\kappa r) \eqno (23) $$
where $H^{(1)}_{m}(x)$ is the Hankel function of the first kind.
The ratio at $r=r_{0}+$ can be expressed as
$$\left. \displaystyle {F^{+}_{jE} \over G^{+}_{jE}}
\right|_{r=r_{0}+}
=-i \left(\displaystyle {M+E \over M-E }\right)^{1/2} \displaystyle
{H^{(1)}_{j-1/2}(i\kappa r_{0}) \over H^{(1)}_{j+1/2}(i\kappa r_{0}) }
\eqno (24) $$
In this case, we can rearrange Eq. (11) as
$$i\left[\frac{M+E}{M-E}\right]^{1/2}\left[\frac{H_{j-1/2}(i\kappa r)}
{H_{j+1/2}(i\kappa r)}\right]=\alpha\left[1-\frac{M+E}{M-E}\cdot
\frac{J_{j-1/2}(i\kappa r)}{J_{j+1/2}(i\kappa r)}\cdot
\frac{H_{j-1/2}(i\kappa r)}{H_{j+1/2}(i\kappa r)}\right]
\eqno(25)$$
from which, we can obtain the energy eigenvalue $E$ with respect to the
different angular momentum $j$.

\vskip 0.5cm
\begin{center}
{\large\bf {4. Concluding remarks}}
\end{center}

In this paper we have carried out the 2+1 Dirac equation with
the $\delta$-potential. It is found that the ratio $F_{jE}/G_{jE}$ of the
radial wave functions $F_{jE}$ and $G_{jE}$ at $r_{0}$ plays an
important role in establishing the second boundary condition (18) from which we can obtain the energy eigenvalue $E$.
This is a very simple method to study the 2+1 Dirac
equation in configuration space.  On the other hand, it is surprisingly found that
the relation between the radial functions $F_{jE}^{+}$, $G_{jE}^{+}$ and $F_{jE}^{-}$, $G_{jE}^{-}$ at the both sides of the potential can be established by $SO(2)$ group, which is a very new point,  to our knowledge.

\vspace{10mm}
{\Large \bf Acknowledgments}  S. H. Dong
would like to thank Prof. A. Frank for
invitation to Universidad Nacional Autonoma de Mexico.
This work is supported by CONACyT, Mexico, under project 32397-E.

\end{document}